\documentclass[aps,twocolumn,amssymb,amsmath,amsfonts,superscriptaddress,nofootinbib]{revtex4-1}

\usepackage{hyperref}
\usepackage{breakurl}

\usepackage{graphicx}	
\usepackage{amsmath}	
\usepackage{amssymb}	
\usepackage{float}      

\usepackage{xcolor}
\hypersetup{
    colorlinks,
    linkcolor={red!50!black},
    citecolor={green!50!black},
    urlcolor={blue!80!black}
}

\def\beq{\begin{equation}}\def\eeq{\end{equation}}
\def\bea{\begin{eqnarray}}\def\eea{\end{eqnarray}}

\newfont{\cursive}{pzcmi at 9pt}

\newcommand{\LONGBIB}[2]{#1}

\newcommand{\AUTHAND}{and }
\newcommand{\ARXIVFULL}[1]{\LONGBIB{ [#1]}{}}
\newcommand{\CITELVC}{\LONGBIB{B.~P.~Abbott {\it et al.} [LIGO Scientific \AUTHAND Virgo Collaborations]}{LIGO Scientific \AUTHAND Virgo Collaborations}}
\newcommand{\CITEDOI}[1]{\LONGBIB{doi:#1}{}}

\begin{document}

\title{Low significance of evidence for black hole echoes in gravitational wave data}

\author{Julian Westerweck}
\email{julian.westerweck@aei.mpg.de}
\affiliation{Max-Planck-Institut f\"ur Gravitationsphysik, D-30167 Hannover, Germany}
\affiliation{Leibniz Universit{\"a}t Hannover, D-30167 Hannover, Germany}

\author{Alex B. Nielsen}
\email{alex.nielsen@aei.mpg.de}
\affiliation{Max-Planck-Institut f\"ur Gravitationsphysik, D-30167 Hannover, Germany}
\affiliation{Leibniz Universit{\"a}t Hannover, D-30167 Hannover, Germany}

\author{Ofek Fischer-Birnholtz}
\email{ofek@mail.rit.edu}
\affiliation{Max-Planck-Institut f\"ur Gravitationsphysik, D-30167 Hannover, Germany}
\affiliation{Leibniz Universit{\"a}t Hannover, D-30167 Hannover, Germany}
\affiliation{Rochester Institute of Technology, Rochester, NY 14623, USA}

\author{\\Miriam Cabero}
\affiliation{Max-Planck-Institut f\"ur Gravitationsphysik, D-30167 Hannover, Germany}
\affiliation{Leibniz Universit{\"a}t Hannover, D-30167 Hannover, Germany}

\author{Collin Capano}
\affiliation{Max-Planck-Institut f\"ur Gravitationsphysik, D-30167 Hannover, Germany}
\affiliation{Leibniz Universit{\"a}t Hannover, D-30167 Hannover, Germany}

\author{Thomas Dent}
\affiliation{Max-Planck-Institut f\"ur Gravitationsphysik, D-30167 Hannover, Germany}
\affiliation{Leibniz Universit{\"a}t Hannover, D-30167 Hannover, Germany}

\author{Badri Krishnan}
\affiliation{Max-Planck-Institut f\"ur Gravitationsphysik, D-30167 Hannover, Germany}
\affiliation{Leibniz Universit{\"a}t Hannover, D-30167 Hannover, Germany}

\author{Grant Meadors}
\affiliation{Max-Planck-Institut f\"ur Gravitationsphysik, D-30167 Hannover, Germany}
\affiliation{Max-Planck-Institut f\"ur Gravitationsphysik, D-14476 Potsdam-Golm, Germany}
\affiliation{OzGrav, School of Physics \& Astronomy, Monash University, Clayton 3800, Victoria, Australia}

\author{Alexander H. Nitz}
\affiliation{Max-Planck-Institut f\"ur Gravitationsphysik, D-30167 Hannover, Germany}
\affiliation{Leibniz Universit{\"a}t Hannover, D-30167 Hannover, Germany}

\begin{abstract}
Recent detections of merging black holes allow observational tests of the nature of these objects.
In some proposed models,
	non-trivial structure at or near the black hole horizon could lead to
	echo signals in gravitational wave data.
Recently, Abedi et al. claimed tentative evidence for repeating damped echo signals
	following the gravitational-wave signals of the binary black hole merger events
	recorded in the first observational period of the Advanced LIGO interferometers.
We reanalyse the same data,
	addressing some of the shortcomings of their method using more background data and a modified procedure.
We find a reduced statistical significance for the claims of evidence for echoes,
	calculating increased p-values for the null hypothesis of echo-free noise.
The reduced significance is entirely consistent with noise, and so
        we conclude that the analysis of Abedi et al. does not provide any observational evidence
        for the existence of Planck-scale structure at black hole horizons.
\end{abstract}

\maketitle

\section{Introduction}

The detections of gravitational wave (GW) signals
	allow for new tests of the nature of black holes
	\cite{Abbott:2016blz,Abbott:2016nmj,TheLIGOScientific:2016pea, Abbott:2017vtc, Abbott:2017gyy, Abbott:2017oio,
			TheLIGOScientific:2016wfe,TheLIGOScientific:2016src}.
Black holes are characterised by their horizons.
In vacuum general relativity these horizons are devoid of material structure.
The possibility that additional structure may form at or near the horizon location
	has been widely discussed in the literature,
	motivated by a number of different models and theoretical considerations \cite{MembraneParadigm, Almheiri:2012rt}.
The Advanced LIGO \cite{Harry:2010zz,TheLIGOScientific:2014jea} and VIRGO \cite{TheVirgo:2014hva}
	detectors have detected gravitational wave signals
	from several binary black hole mergers
	\cite{Abbott:2016blz, Abbott:2016nmj, TheLIGOScientific:2016pea, Abbott:2017vtc, Abbott:2017gyy, Abbott:2017oio}.
These detections now make those ideas testable in the observational regime. 

A generic set of models called Ultra Compact Objects (UCOs)
	\cite{Cardoso:2016oxy, Cardoso:2016rao, Cardoso:2017njb, Mark:2017dnq, Volkel:2017kfj}
	can mimic black holes in terms of their gravitational wave emission at early stages of binary inspirals.
These models are designed to match the properties of standard black holes at sufficiently large distances,
	but differ in the near-horizon regime.
The gravitational wave signal from the inspiral of two UCOs
	is expected to be almost identical to that of standard black holes
	(for possible tidal modifications see \cite{Cardoso:2017cfl}).
However, the merger and ringdown signals may differ sufficiently to be detectable.
Near-horizon material structures motivated by semi-classical and quantum gravity theories could,
        at least partially, reflect incoming waves
	which in standard vacuum general relativity would be fully absorbed by the black hole.

Recent works by Abedi, Dykaar and Afshordi (ADA) \cite{Abedi:2016hgu, Abedi:2017v2, Abedi:2017isz}
	have claimed to find tentative evidence of near-horizon Planck-scale structure using
	data \cite{LOSC, Vallisneri:2014vxa} from the three Advanced LIGO events
	GW150914, LVT151012 and GW151226. 
In the simplified analysis of \cite{Abedi:2016hgu,Abedi:2017v2},
this near-horizon structure gives rise to
so-called echoes \cite{Cardoso:2016oxy, Cardoso:2016rao,Nakamura:2016gri,Holdom:2016nek}.

The data used by ADA is from the
	LIGO Open Science Center (LOSC) \cite{LOSC, Vallisneri:2014vxa},
	which contains a total of 4096 seconds of
	strain data from both Advanced LIGO detectors
	around each of the three events.
Out of this data ADA used only 32 seconds
	centered around each event
	for their analysis.
The authors claimed in \cite{Abedi:2016hgu}
	to find evidence for such echoes
in data following the three events with
a p-value $3.7\times 10^{-3}$,
corresponding to a combined significance of 2.9$\sigma$
(with the one-sided significance convention
used in \cite{Abbott:2016blz,Abbott:2016nmj,TheLIGOScientific:2016pea, Connaughton:2016umz},
this value corresponds to 2.7$\sigma$).
This was subsequently updated to
a p-value of $\sim 1\%$
and interpreted as 2.5$\sigma$-level tentative evidence in \cite{Abedi:2017v2}.
Nonetheless if such a signal were shown to be present in the data,
it would force a major re-evaluation
of the standard picture of black holes
in vacuum Einstein gravity.

Here we investigate concerns about the methods in \cite{Abedi:2016hgu}
	and ADA's updated works \cite{Abedi:2017isz, Abedi:2017v2},
	and give a different significance estimate for the findings.
Our initial caveats concerning \cite{Abedi:2016hgu} appeared as \cite{EchoComments}.
We do not examine the theoretical motivations for the existence of such near-horizon Planck-scale structure,
	nor the model templates for which ADA have chosen to search.
Rather, we focus on the data analysis methods as reported and on the significance estimates assigned to the results.
We identify a number of shortcomings in the analysis and perform an improved analysis,
	which corrects for several of these problems.
We evaluate the echo findings in the gravitational wave data \cite{LOSC, Vallisneri:2014vxa},
    estimate their significance with updated p-values
    (for a general critique of p-values, see \cite{pvalues})
	and conclude that there is as of yet no evidence
	for the existence of black hole echoes in this data.

\section{ADA's Model and search procedure}

\begin{figure}
	\includegraphics[width=\columnwidth]{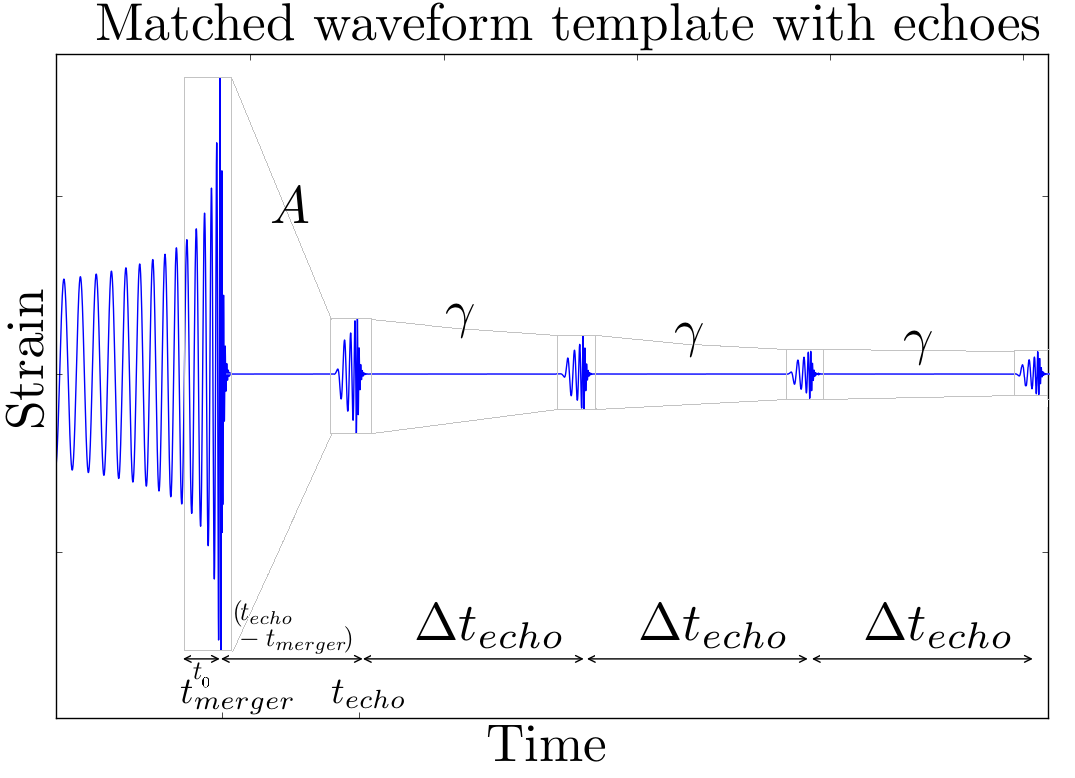}
    \caption{A coalescence template extended to include echoes. The five parameters of the echo waveform model are illustrated, and the phase-inversion between echoes is visible.}
    \label{fig:templates}
\end{figure}
The analysis of ADA \cite{Abedi:2016hgu, Abedi:2017v2, Abedi:2017isz} consists of three parts:
	a simple waveform model,
	a search procedure,
	and a significance estimation method.
In this section we briefly review these.

With a partially reflective surface outside the horizon,
	echo signals may be found as wave solutions in a cavity formed by the near-horizon membrane barrier and 
	the angular momentum barrier (``photon sphere") that exists further out \cite{Cardoso:2016oxy, Cardoso:2016rao}.
In the geometric wave picture,
	at each barrier the wave is partially reflected and partially transmitted.
Each partially transmitted wave from the outer angular momentum barrier would be detected by distant observers as an echo.
The delay time between subsequent echoes results from the travel time between the two barriers.
This time may be different for the first echo due to non-linear effects during the merger,
	as may further parameters of the echo signal such as the damping between successive echoes.
For a description of the echoes as poles of the propagators see \cite{Mark:2017dnq}.

An example of such an echo template is shown in Fig.\ref{fig:templates} and several parameters define its features: 
\begin{enumerate}
    \item $\Delta t_{\rm echo}$: The delay time between subsequent echoes,
            resulting from the travel time between the barriers.
        $\Delta t_{\rm echo, theory}$ is the expected value found in \cite{Abedi:2016hgu,Abedi:2017v2},
            based on the inferred final mass and spin parameters for each event
			\cite{Abbott:2016blz, Abbott:2016nmj, TheLIGOScientific:2016pea, Abbott:2017vtc, AbediPrivate}.
        In the search, the parameter $\Delta t_{\rm echo}$ is allowed to vary around the theoretical value $\Delta t_{\rm echo, theory}$ to account for uncertainties.
    \item $t_{\rm echo}$: The time of the first echo.
        This is expected to be $t_{\rm merger} + \Delta t_{\rm echo}$,
			where $t_{\rm merger}$ is the time of the merger.
        It is allowed to deviate from this expectation in the search to account for non-linear effects close to the merger \cite{Abedi:2016hgu}.
    \item $A$: The amplitude of the first echo relative to the original signal amplitude. 
    \item $\gamma$: The relative amplitude between subsequent echoes. 
    \item $t_0$: Only the last part of the original waveform is used to produce the echo waveform;
            this parameter describes how far before $t_{\rm merger}$ the original waveform is tapered down to 0, using the tapering function given in \cite{Abedi:2016hgu}.
\end{enumerate}

In addition, the phase is inverted between subsequent echoes.
Likewise, the phase-difference between the original signal and the first echo is fixed to $\Delta \phi = \pi$.
We use an abbreviated notation for the combination of parameters
	$x := (t_{\rm echo} - t_{\rm merger} ) / \Delta t_{\rm echo}$,
	with an expected value for the first echo of $x=1$.

The \textbf{ADA-search} procedure used in \cite{Abedi:2016hgu,Abedi:2017v2} consists of the following steps:
\begin{enumerate}
    \item Produce a pure echo template for given echo-parameters.
        The original event is removed from the template.
    \item Produce a bank of these templates, with an evenly spaced grid in the parameters listed above. 
    \item Perform matched filtering with the echo templates. The original event is removed from the data prior to this. 
    \item Maximise SNR$^2$ over all parameters for each value of $x$. 
\end{enumerate}

The maximisation uses either each single event or combinations of events.
The combination assumes some parameters to be different between events,
    namely $A$ and $\Delta t_{\rm echo}$.
The parameters $x$, $t_0/\Delta t_{\rm echo, theory}$ and $\gamma$ are kept identical for each event. 
For combinations of events, the sum of the individual SNR$^2$s is maximised. 

The \textbf{ADA-estimation} uses the following method to estimate the significance of their findings \cite{Abedi:2016hgu}:
\begin{enumerate}
    \item Find the maximum SNR$^2$ value in the range $x \in (0.99,1.01)$ after the event. 
    \item Calculate and maximise SNR$^2$ over the time range
			$9 \leq \frac{t_{\rm echo} - t_{\rm merger}}{\Delta t_{\rm echo, theory}} \leq 38 $.
        The maximisation is slightly adapted for this step. 
    \item Divide this last range into 1450
		segments, 	each of duration $2\%$ of
		$\frac{t_{\rm echo} - t_{\rm merger}}{\Delta t_{\rm echo, theory}}$.

    \item A p-value is found as the number of segments with higher peak SNR$^2$,
        divided by the total number of segments.
\end{enumerate}

\section{General remarks}

A first immediate problem arises regarding
	how strong the relative signal should be for the three events.
The two binary black hole events GW150914 and GW151226 were detected by the Advanced LIGO detectors
	with significance levels $>5.3\sigma$
	and signal-to-noise ratios of $23.7$ and $13.0$
	respectively \cite{TheLIGOScientific:2016pea}.
The other event, LVT151012,
	had a reported significance of only $1.7\sigma$ and a signal-to-noise ratio of $9.7$
	combined between the two Advanced LIGO detectors.
However, in Table II of \cite{Abedi:2017v2} we see that the signal-to-noise ratio
	of the claimed echo signal is actually largest for LVT151012.
	
The higher SNR of LVT151012 cannot be due to
	the different projected number of echoes between the events.
The different $\Delta t_{\rm echo}$ leads to differing numbers of echoes in a given duration:
	the 32 seconds of data used would contain ($\sim\!180$) for LVT151012
	and ($\sim\!110$) for GW150914.
Although the number of echoes is larger for LVT151012,
	late echoes are strongly damped.
They decrease by a factor of 10 over $\sim\!22$ echoes
	for the claimed relative amplitude $\gamma\sim0.9$.
Thus in order for the echoes of LVT15012
	to have a higher SNR than the echoes of GW150914, 
	their amplitude must be very high.
In fact to account for the reported SNRs,
	the initial amplitude for the first echo of LVT151012
	would have to be about $10\%$ higher than that of GW150914 \cite{EchoComments},
	while the original event's peak is about 2-3 times lower for LVT151012 in comparison to GW150914's.
This would require their parameter $A$ to be about 2-3 times larger for LVT151012 than for GW150914.
This seems to be confirmed by the best fit search results in Table II of the updated work \cite{Abedi:2017v2},
	which gives $A_{\rm GW150914} = 0.091$ and $A_{\rm LVT151012} = 0.34$.

We assume that far in the wave zone the gravitational wave signal of the echoes decays similarly to the signal of the event itself,
    i.e. linearly with the distance from the source.
This explicit astrophysical assumption, in addition to those in \cite{Abedi:2016hgu, Abedi:2017isz, Abedi:2017v2},
    is the basis for the above concern.
The lower significance of LVT151012 is rooted in its distance:
    its mean estimated distance being more than twice as large as that of GW150914 and GW151226, we expect weaker echo signals.
While particular combinations of system parameters and signal morphologies may have significant effects on the generation of echoes and their relative amplitudes,
    changing their relative significance, there is yet no extensive model to justify abandoning this concern here.
    
The inferred amplitude parameters suggest that
a lot of gravitational wave energy was emitted in the echoes:
a very rough calculation implies that the amount
of energy emitted in the echoes was approximately 0.1 solar masses (for GW150914)
and 0.2 solar masses (for LVT151012).
This should be compared to the total estimated energy emitted by the original signal
of 3 solar masses (for GW150914) and 1.5 solar masses (for LVT151012).

We also note an inconsistency in the above procedure, 
    resulting from the use of a fixed waveform for each event as the basis for all echo templates, obtained from the LOSC \cite{LOSC}. 
The parameters of the echo templates, in particular $\Delta t_{\text{echo}}$, depend on the mass and spin parameters of the final black hole.
Instead of using only one initial waveform and generating all echo templates with this, one should use an initial waveform that corresponds to each set of echo parameters to be varied over. 
Using the single LOSC waveform is a simplification, restricting to only one choice of final mass and spin parameters for the echoed original event, while simultaneously varying over the final mass and spin values through $\Delta t_{\text{echo}}$. 

\section{Validation of the matched-filter analysis}

\begin{figure}
	\includegraphics[width=\columnwidth]{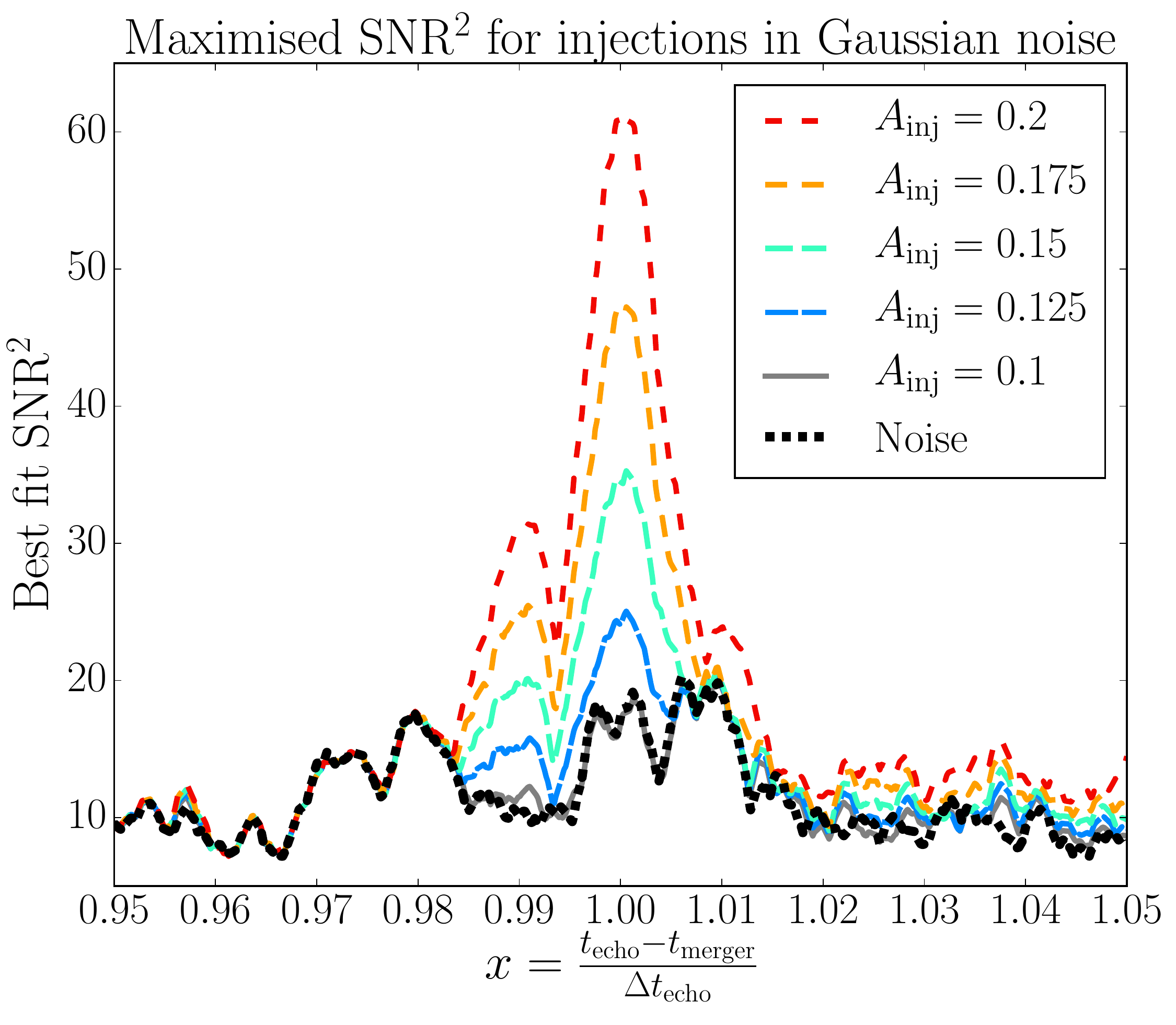}
    \caption{The matched filtering technique is able to recover signals with a variety of amplitudes.
			As shown here, the SNR depends on the amplitude of the signal.
		The amplitude found by ADA ($A=0.1$) is close to the level that is found in pure Gaussian noise.
		An amplitude twice as large as this would be clearly identifiable in the data.
}
    \label{fig:recovery}
\end{figure}

We wrote a separate implementation of the \textbf{ADA-search} procedure,
    that we refer to as \textbf{ADA$_{{AEI}}$-search}.
No changes were made to the algorithm as described before,
    while the implementation itself is independent.
The SNR$^2$-results obtained with our implementation are similar to those shown in \cite{Abedi:2016hgu}. 

As a first check, we verify that the \textbf{ADA$_{{AEI}}$-search} procedure can distinguish between pure noise and simulated echo-signals.
For this, a known signal is injected into simulated noise.
We simulate Gaussian noise
    with a Power Spectral Density (PSD) similar to that found for the detector data around each event
    (calculated from the LOSC data).
The \textbf{ADA$_{{AEI}}$-search} is then applied to simulated data
    of both pure noise and also the same noise with added injections of different amplitudes.
In this test, we only use echo waveforms with parameters similar to the best-fit results of \cite{Abedi:2016hgu,Abedi:2017v2}.
Fig. \ref{fig:recovery} shows the dependence of the SNR$^2$ peak on the injection amplitude $A_{\rm inj}$.
The effectiveness of the method in finding a signal depends on $A_{\rm inj}$.
This test was performed for different realisations of the simulated noise.
The minimum $A_{\rm inj}$ required to find a peak rising above the noise background also depends on the noise instantiation.
We find that $A_{\rm inj} \sim 0.1$ can yield a visible peak.
This is the best-fit value of $A$ reported for GW150914 in \cite{Abedi:2016hgu}.
In one out of the five trials conducted in this first test,
    however, a higher amplitude was necessary to distinguish the signal from noise,
    as shown in Fig. \ref{fig:recovery},
    where the noise and the quietest injection have almost identical SNR$^2$ results.
This prompted us to perform more detailed statistical analyses and injection-recovery analyses, as described below.

\section{Prior ranges and template spacing}

Values for each echo parameter are determined from within a prior range.
Each template in the bank is produced for a specific value of each parameter.
The matched filtering method finds a higher SNR for data similar to the template,
    but each template can recover signals with a range of parameter values. 
The three parameters $\gamma$, $t_0$ and $t_{\rm echo}$
    are determined by maximisation,
    with $\gamma$ and $t_0$ kept fixed between the different events.
In this, the parameters recovered are defined as the values corresponding to
    the template in the bank which yields the highest SNR. 
The maximisation is performed over all templates in the bank 
    and thus over all values in the parameter grid used to create the bank.
The boundaries of the parameter grid are determined by a prior range,
    where the ranges chosen by ADA are displayed in Table I of \cite{Abedi:2016hgu}. 

\begin{figure}
  \includegraphics[width=\columnwidth]{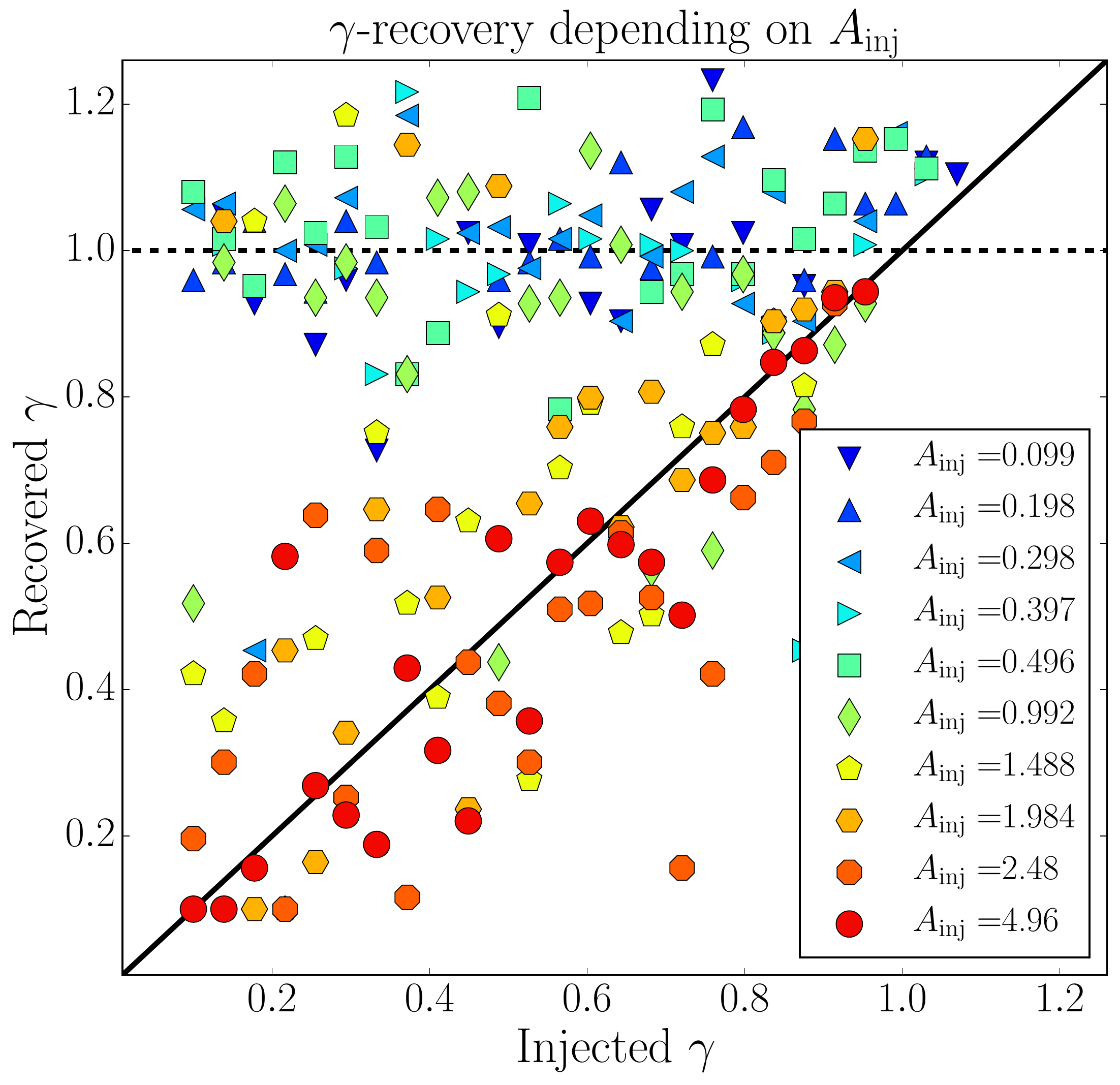}
  \caption{Injected and recovered values for $\gamma$, correct recovery would be seen as a diagonal.
                        The search method's preference for $\gamma=1$ (dashed line) at lower injection amplitudes is clearly seen.}
  \label{fig:inj_rec_gamma}
\end{figure}

\begin{figure}
  \includegraphics[width = \columnwidth]{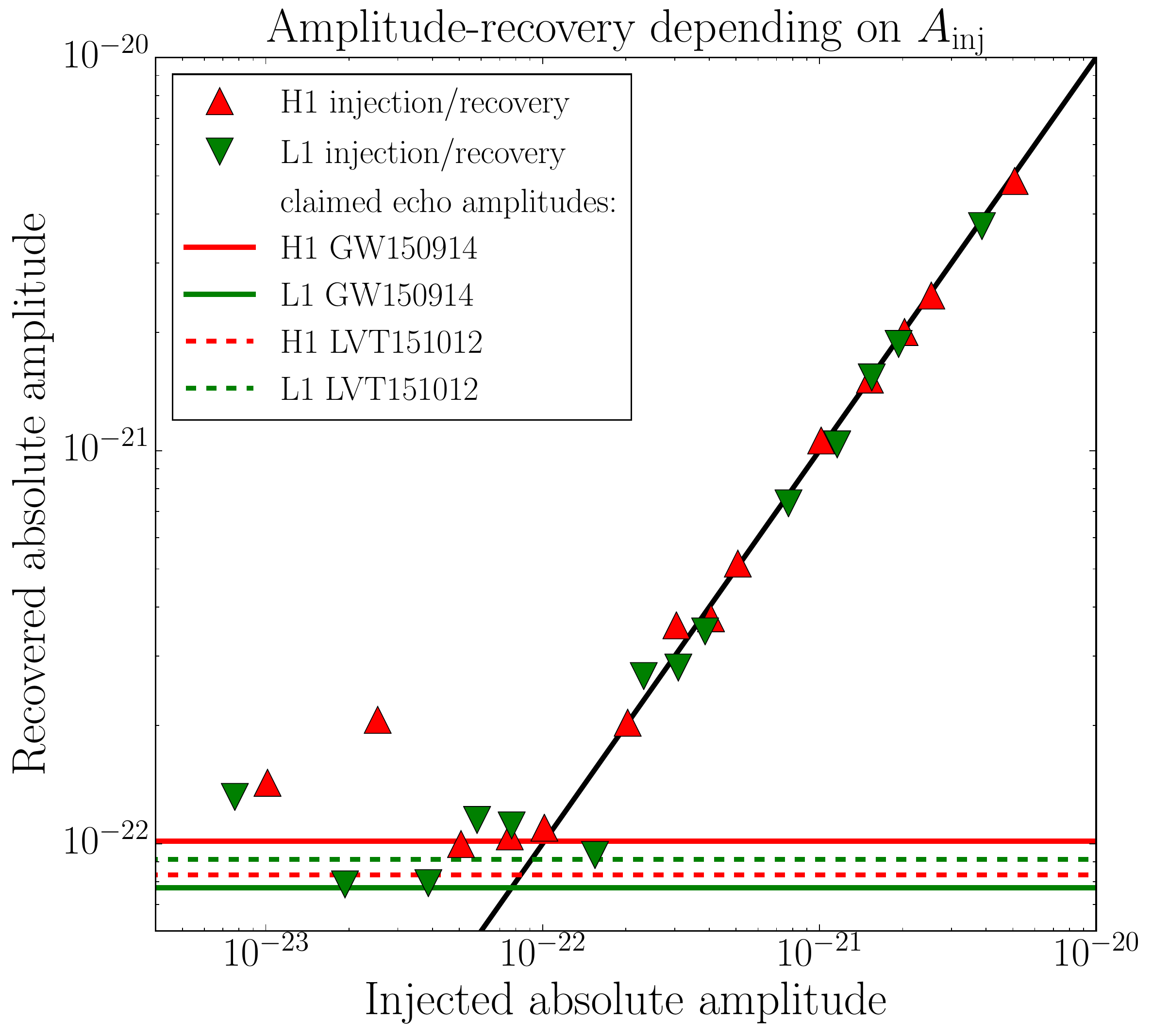}
  \caption{Injections of echo signals into Gaussian noise are analysed for different amplitudes of the injection.
            For injections with amplitudes above $10^{-22}$ of peak strain,
                the recovered values are close to the injected ones,
                indicated by the diagonal line.
            For injections with lower amplitudes, the recovered amplitudes are around $10^{-22}$,
                independently of the injected value.
            The amplitudes found in \cite{Abedi:2016hgu} are shown as horizontal lines (``true'' values unknown)
                and are similar to the values incorrectly recovered for low amplitudes.
            The shown injections are made with $\gamma=0.8$, with similar results for other values.}
  \label{fig:amplitude_recovery}
\end{figure}

The values for $\gamma$ and $t_{0}$ resulting from this maximisation
    are found to lie very close to the boundary of their prior
    range, $0.9$ and $-0.1$ respectively \cite{EchoComments}.
This suggests that there may be support
for values of these parameters that lie outside of this range.
If these values reflect the priors rather than the data, 
    then they cannot be reliably considered as evidence for a detection claim. 
Furthermore, a value greater than unity for $\gamma$ means that
    each successive echo has an amplitude greater than the previous echo.
Such a result would require the echo signal to be extracting energy from the black hole spacetime.

We tested whether the preference for these parameter values is an artifact of the method, 
    again using known signals injected into simulated noise. 
We constructed Gaussian noise with a PSD estimated from the 4096 seconds of LOSC data around GW150914. 
The injected signals are pure echo signals based on the LOSC GW150914-template for various echo parameters. 
The range of $\gamma$ is widened to $\gamma \in (0.1,2.0)$ both in the prior of the search and the injections.
The range of $t_0$ is widened to $t_0 \in (-0.2,0) \Delta t_{\rm echo, theory}$ in the search.
It is not widened for the injections in this test, as the dependency of the maximised SNR on the wider range in $t_0$ was found to be much weaker than for $\gamma$. 
The relative amplitude of the injections $A = A_{\rm inj}$ ranges from ADA's recovered value $0.1$ to about 50 times this amplitude.
We then compare the best-fit value of $\gamma$ from the search with the value of the injection. 
This comparison is shown in Fig. \ref{fig:inj_rec_gamma}.

The \textbf{ADA-search} method is biased towards finding $\gamma$ values close to $1$. 
Ideally, the recovered parameter value would be closest in the grid to the value of the injection.
In Fig. \ref{fig:inj_rec_gamma}, this would mean lying as close as possible to the plotted diagonal.
In this figure, the recovered values are close to the injected ones for higher injection amplitudes.
Thus for very high echo amplitudes, the recovery method could in principle be effective.
For lower injection amplitudes, there is a preference for recovered values of $\gamma$ close to $1$,
    independently of the $\gamma$ value of the injection.
Thus finding $\gamma \sim 1$ as the best-fit value in the search does not necessarily mean that this is indeed the correct value for an existing signal.
The method is biased to find these values for $\gamma$ in almost all cases.
In particular this is also true for relatively low echo amplitudes as found by ADA, and even significantly higher signal amplitudes.
We interpret the recovery of $\gamma \sim 1$ as a generic property of the method and finding such a value cannot be considered evidence for the presence of a signal. 

\begin{figure}
  \includegraphics[width = \columnwidth]{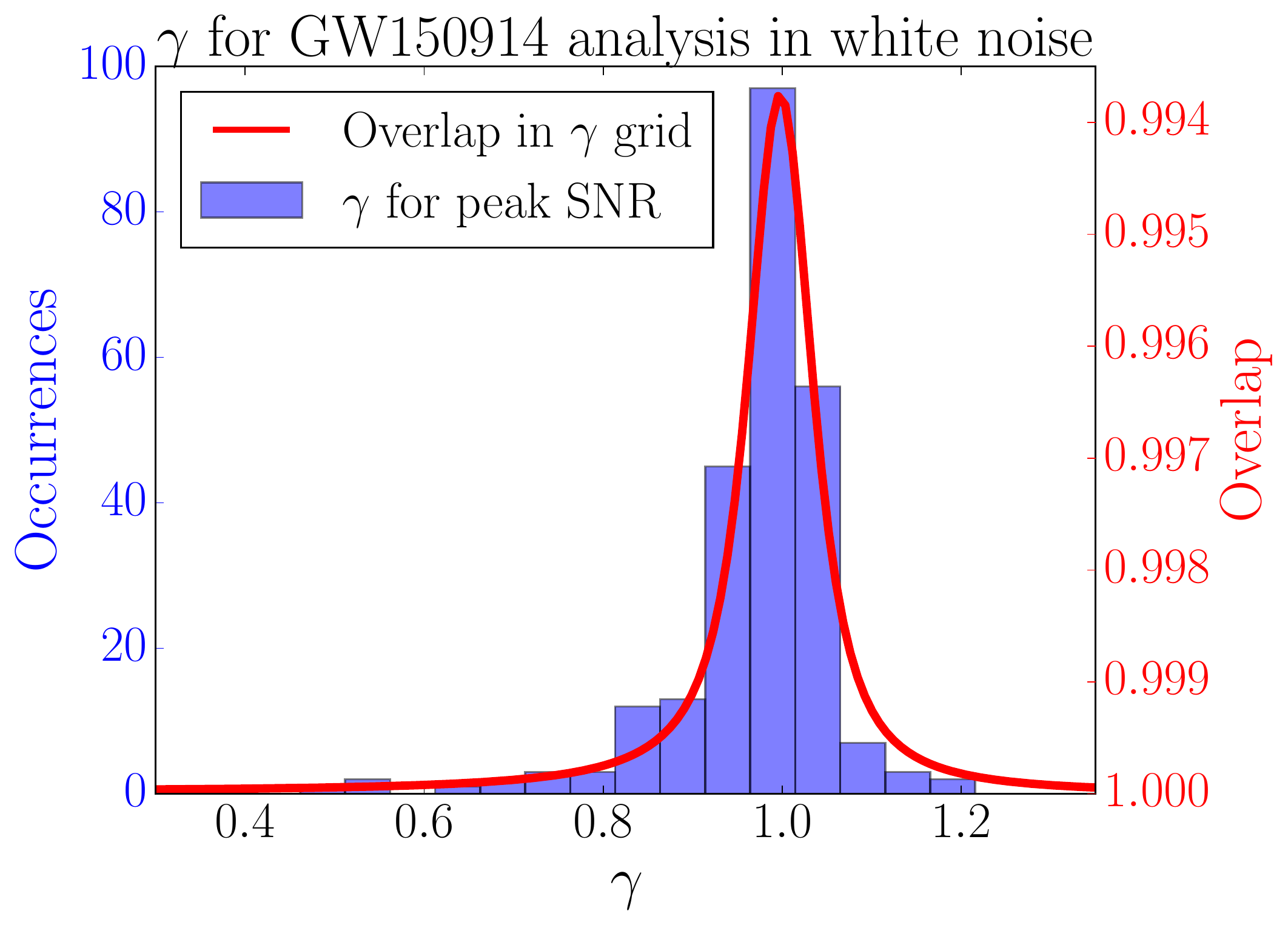}
  \caption{Demonstration of the preference for $\gamma=1$ in white Gaussian noise
        and the correspondence of the recovered distribution with the spacing in the template grid.}
  \label{fig:gamma_white_noise}
\end{figure}

The bias is introduced through the spacing between templates in the bank,
    as can be found through an analysis on white Gaussian noise and calculating the overlap between the templates. 
An analysis in white Gaussian noise using the same parameter range as ADA also
    shows a strong preference for $\gamma = 0.9$.
Further extending the range to $\gamma \in \left( 0.1, 2.0 \right)$
    displays preference for $\gamma=1$ in white noise.
The distribution of recovered $\gamma$ values in this test is shown in Fig. \ref{fig:gamma_white_noise}.
The reason for this is revealed by calculating the overlap between neighbouring templates 
    in the parameter grid for different $\gamma$, while keeping the other parameters fixed.
As can be seen in Fig. \ref{fig:gamma_white_noise},
    the match between neighbouring templates follows the same distribution as the preference for recovering $\gamma=1$.
Templates with $\gamma$ close to $1$ lie further apart in the noise weighted match metric
    than other templates.
Each template near $\gamma=1$ covers a larger region
    of the signal space than other templates and thus,
    more noise realisations are best matched by the (morphology-wise) more scarcely placed templates close to $\gamma=1$.

We similarly test how the method recovers the echo signal's amplitude through injections into Gaussian noise\footnote{we thank N. Afshordi for suggesting this additional test\cite{Abedi:2018pst}}.
For the results in Fig. \ref{fig:amplitude_recovery},
    we chose to show the absolute peak amplitude of the echo signal instead of the parameter $A$,
    which gives the echo amplitude relative to the original event amplitude.
This allows us to find the minimum echo amplitude to be recovered correctly,
    independently of that of the event.
Fig. \ref{fig:amplitude_recovery} shows the recovered values deviate from the injected ones strongly below strain amplitudes of about $10^{-22}$.
For lower injection amplitudes, values around $10^{-22}$ are found instead of the injected ones.
This suggests that finding such low amplitude values might be expected in pure noise as well.
The absolute value is close to those found in \cite{Abedi:2016hgu}:
    the relative amplitudes of $0.1$ for GW150914 and $0.3$ for LVT151012, multiplied by the respective events' peak amplitudes,
    are shown as horizontal lines in Fig. \ref{fig:amplitude_recovery}, and seem consistent with incorrect recovery of the method for lower injection amplitudes. 

Extending the template bank to a wider range in $\gamma$ and $t_0$
    and performing the same analysis as before leads to a modified SNR structure in $x$,
    where additional and higher peaks appear further away from the predicted value for the echoes in GW150914. 
As we will see below,
    applying a wider parameter range also for the background estimation
    results in a further factor $\sim 3.5$ increase in the p-value of the combination (1,2,3)
    (using the 32-second dataset estimation).
The modified p-values for the wider priors of different combinations are found in Table \ref{tab:16+32_original+widened},
    where the widened prior entries refer again to the ranges $\gamma \in (0.1,2.0)$ and $t_0 \in (-0.2,0) \Delta t_{\rm echo, theory}$.

\section{Extending the background estimation}

To calculate a significance for the match found in the templated search,
	an assessment of the background must be performed.
Since an analytical noise model is not known, 
    real data away from the possible signal is used to estimate the noise background. 
This relies on an assumption that the background noise is similar to that during the time of interest.
The noise background is calculated by how often an equal or larger SNR value is obtained in the off-source data.
ADA chose to do this in a short period of time of approximately 16 seconds of data after each event.
To obtain sufficient background statistics this period of time was used intensively: 
    they consider 16 second stretches of data as independent when shifted by only 0.1 seconds.

This background estimation is problematic \cite{EchoComments} for two reasons: 
    potential contamination of the background samples by existing echo signals,
    and the lack of independence between background samples. 
The estimation uses a range of $t_{\rm echo}$ values that is only $\mathcal{O}(10)$ 
	echo periods away from the merger. 
If there is indeed an echo signal in the data then this region
	will not be entirely free of the signal being searched for.
At the beginning of the region the amplitude of the echoes
	would only have dropped by a factor $0.9^9 \!\sim\! 0.4$.
One therefore expects a contaminated
background estimation. Even in the absence of echoes, a
random feature mistaken for echoes in one segment may
well extend to neighbouring segments, and they cannot be treated as independent (see discussion of template
auto-correlations below for the problem of insufficient independence of samples).

Each of the data sets released at the LOSC \cite{LOSC} consists of 4096 seconds of data.
Both GW150914 and LVT151012 are located 2048 seconds into this data,
	equivalent to $\mathcal{O}(10^3)$ echo periods,
	thus for large stretches of the data, such contamination through the presence of a damped echo signal would be negligibly small. 

\begin{figure}[H]
	\includegraphics[width=\columnwidth]{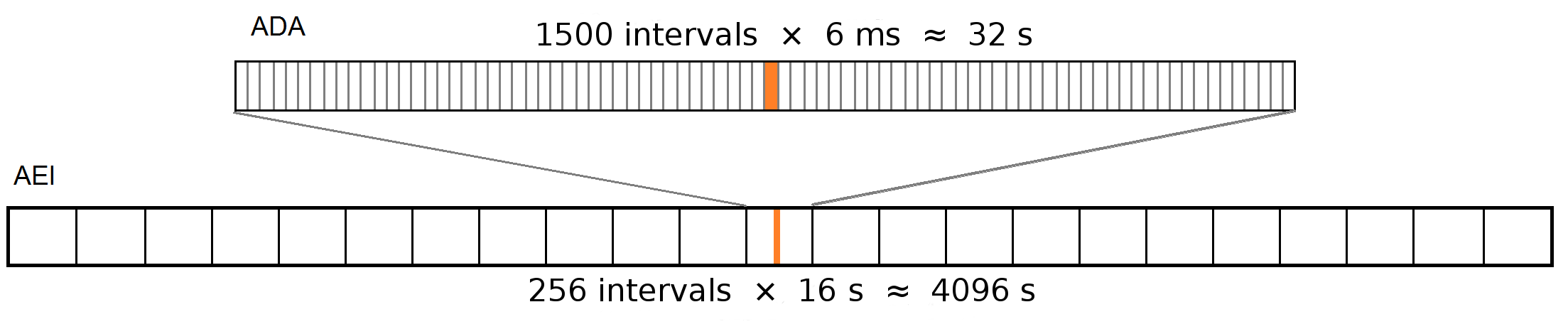}
    \caption{Schematic comparison of the data segments used to estimate the background between
				ADA \cite{Abedi:2016hgu,Abedi:2017v2} and this work (AEI).
				Compared to ADA, we extend the amount of data used for background estimation
				to the full 4096 seconds for each event available from the LOSC \cite{LOSCtutorial}.}
    \label{fig:wider}
\end{figure}

We have performed a different background estimation as an independent test, 
    which uses the full period of 4096 seconds of LOSC data available around each event.
A schematic comparison of the different choices of data used for background estimation is shown in Fig. \ref{fig:wider}.

\begin{figure}[H]
    \includegraphics[width=\columnwidth]{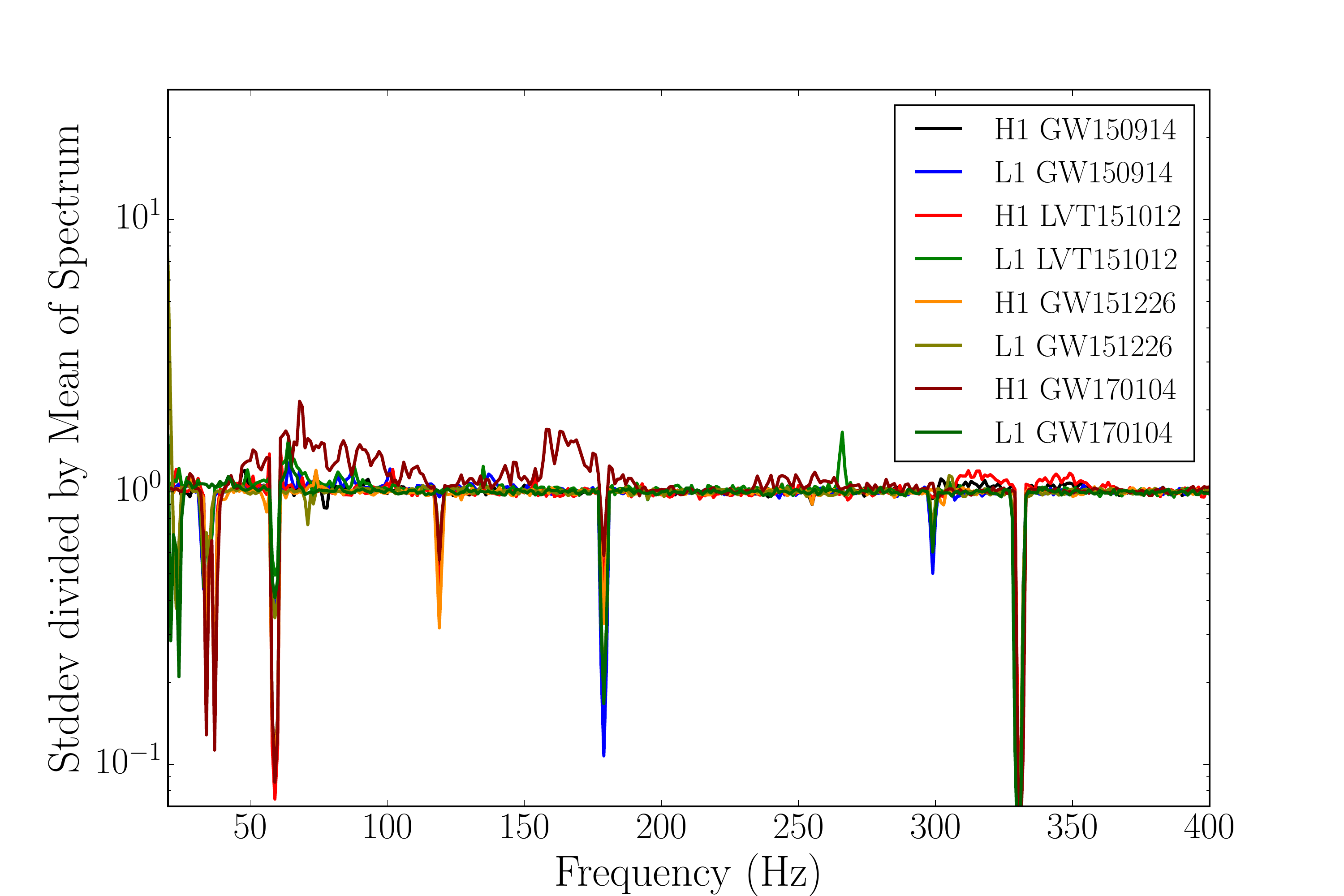}
    \caption{Rayleigh plot of noise variation,
                showing the ratio of the standard deviation to the mean for frequency bins of the PSD
                estimated using 16 second segments of the 4096 second data stretch for both detectors for each event. 
                }
    \label{fig:noiserayleigh}
\end{figure}

\begin{figure}[H]
	\includegraphics[width=\columnwidth]{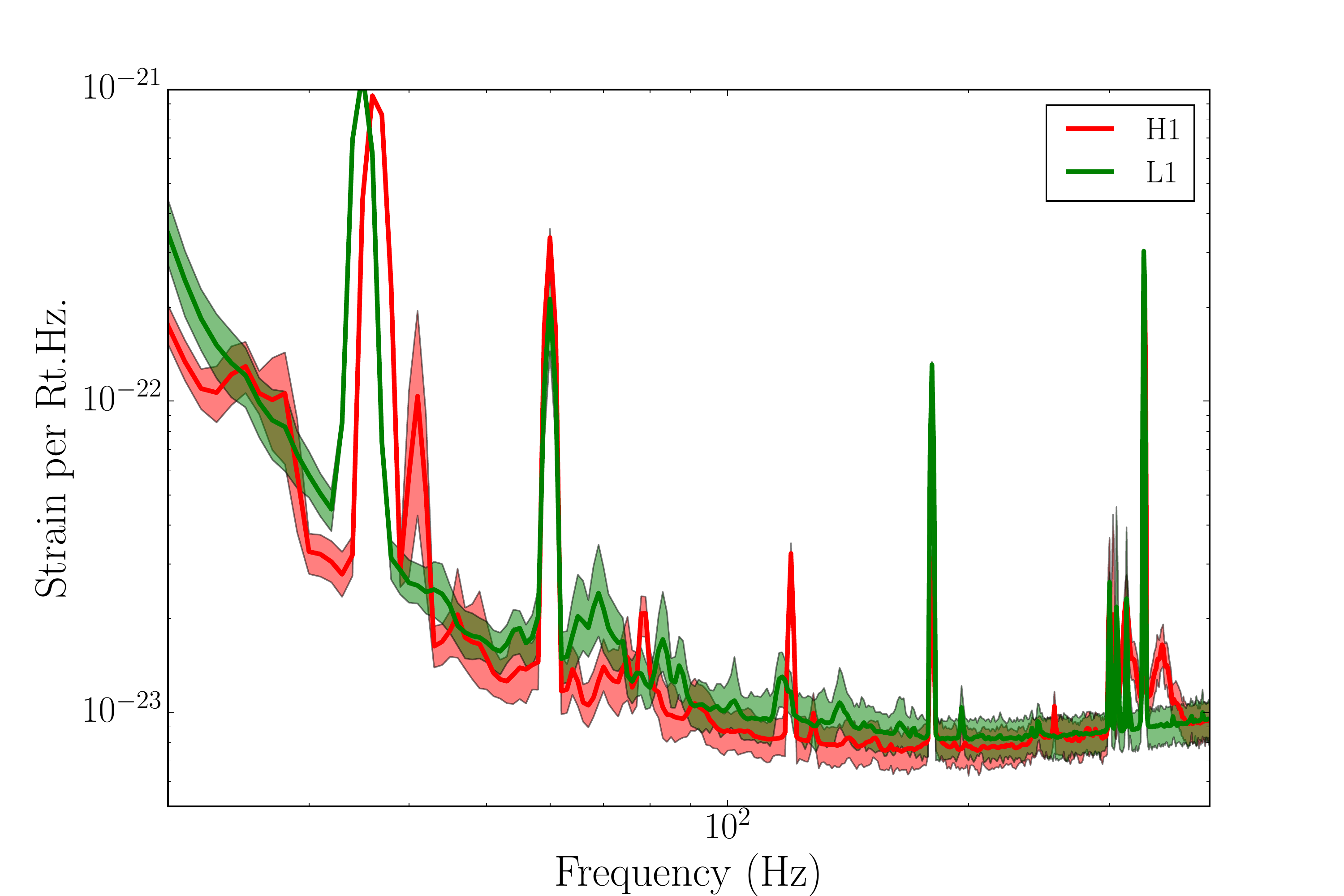}
    \caption{Variation of the noise spectrum during the 4096s around GW150914
				calculated using 16 second segments,
				showing the $1\%$ and $99\%$ percentile variations.
			In this sense, the data is sufficiently stationary for background estimation to be reliable
				during the full 4096 second data stretch.}
    \label{fig:noise}
\end{figure}

The obtained p-value and background estimate are only meaningful
	if the data in the background is comparable to that at the time of the event.
A plot of the noise variations over the full 4096 seconds of data released for each event
	is shown in Fig. \ref{fig:noiserayleigh} and for GW150914 specifically in Fig. \ref{fig:noise}.
The variations are seen to be small and we conclude that for the four events considered our background estimate
	is indeed characteristic of the noise just after the event.
For the graph showing the properties of noise in the Hanford detector around the time of LVT151012 in Fig. \ref{fig:noiserayleigh}, a reduced amount of data was used. 
This choice is made due to three loud short transient noise features, which we discuss further below. The noise features strongly influence the Rayleigh statistic calculation, while occupying less than 0.1 \% of the data. 
Using data excluding these noise features, the variation as shown in Fig. \ref{fig:noiserayleigh} is found. 
Properties of the data at and around LVT151012 are discussed in \cite{DetChar}.

In our case, the 4096 seconds of data for each event are divided into 128 independent,
        32-seconds long segments.
For each, the echoes analysis is performed as it was on the 32-second segment containing the event.
The resulting peak SNR in $x \in (0.99,1.01)$ is found for each segment.
Simply counting the number of segments containing a higher peak SNR
	in this interval yields an estimate for the p-value. For the combined first three events, 
	GW150914, LVT151012 and GW151226,
	our resulting p-value of $0.032$ is about a factor $3$
	larger than the value of $0.011$ found in \cite{Abedi:2017v2},
	where less data and overlapping intervals were used. 

\newpage
An estimate of the p-value significance in this way is susceptible to small number statistics
(accounting for the Poisson error as suggested in \cite{Abedi:2018pst},
    the p-value can be $0.032\pm 0.016$,
    still larger than in \cite{Abedi:2016hgu, Abedi:2017v2}).
The original LOSC templates, before introducing echoes, 
	contain an approximately 16 seconds long waveform
	followed by 16 seconds of a flat zero template. 
	Echoes were introduced only into this flat region
	when producing the echo templates.
After removing the original event, we are left with a 32-second template with 16 seconds of no signal,
	followed by the produced echoes.
So we can double the number of background samples without losing independence between samples,
    by dividing the available data into 256 independent segments of 16 seconds length. 

The exact number of available data segments varies slightly for each of the events.
This is due to the positions of the original signals, and the influence of one of the three mentioned short transient noise features
	(inconsistent with the echo morphology) in one segment of the LVT151012 data, which was discarded.
This short noise feature was found by noticing a very high SNR outlier for one data segment.
The feature can be seen in the whitened time-domain data, appearing close to the beginning of the data segment. 
The search procedure always aligns one of the first and thus loudest echoes with the noise feature,
    yielding the high SNR.
The effect of not discarding the high SNR noise dataset always is an increase of p-value due to the very high SNR,
    but the effect on the resulting p-values is minimal ($\sim 1/256$).
Only the estimation with 16-second long segments is influenced by this noise feature. 
The total number of estimates when combining events is thus 125 to 126 for the 32-second segments
    and 250 to 251 for the 16-second segments. 

The other two short noise features appear late in the respective data segments. 
For these, the search does not consistently align one of the later and more strongly damped echoes with the noise feature, as the increase in SNR is outweighed by the placement of the first loud echoes in the data. Thus the search is not influenced by these features significantly and we do not exclude the data segments from the estimation. 

The results of this alternative approach for the significance estimation,
    both using 32- and 16-second long segments, are shown in Table \ref{tab:16+32_original+widened}.
Different combinations of the events are considered,
    denoted chronologically as
    (GW150914, LVT151012, GW151226) $\rightarrow (1,2,3)$.
In addition, for comparison and as the first detection after the claims of \cite{Abedi:2016hgu},
    we also consider the first event in the second observing run, GW170104 \cite{Abbott:2017vtc}, denoted $4$ in Table \ref{tab:16+32_original+widened}.

For the combination $(1,2,3)$, a p-value of $0.011$ was found in \cite{Abedi:2016hgu}.
With four points out of 125 trials giving maximised combined SNR values larger than immediately after each event,
	our method finds the p-value for the particular SNR value to be $4/125 = 0.032$ thus $\sim 3\%$.
Using twice the number of samples of 16 seconds length each, we find $5/250 \sim 2 \%$ for the p-value.

To highlight the role of LVT151012 in obtaining low p-values,
    we have chosen to make a comparison with a combination of three events excluding LVT151012.
When choosing the available events to be combined in the analysis,
	a reasonable choice seems to be using those of sufficiently high significance.
Here, this means GW150914, GW151226 and GW170104,
	the combination $(1,3,4)$, for which we find $9/125 \sim 7\%$ and $50/251 \sim 20\%$ respectively. These values are much higher than for combinations including LVT151012 and are fully consistent with the pure noise null hypothesis.

\begin{figure}
  \includegraphics[width=\columnwidth]{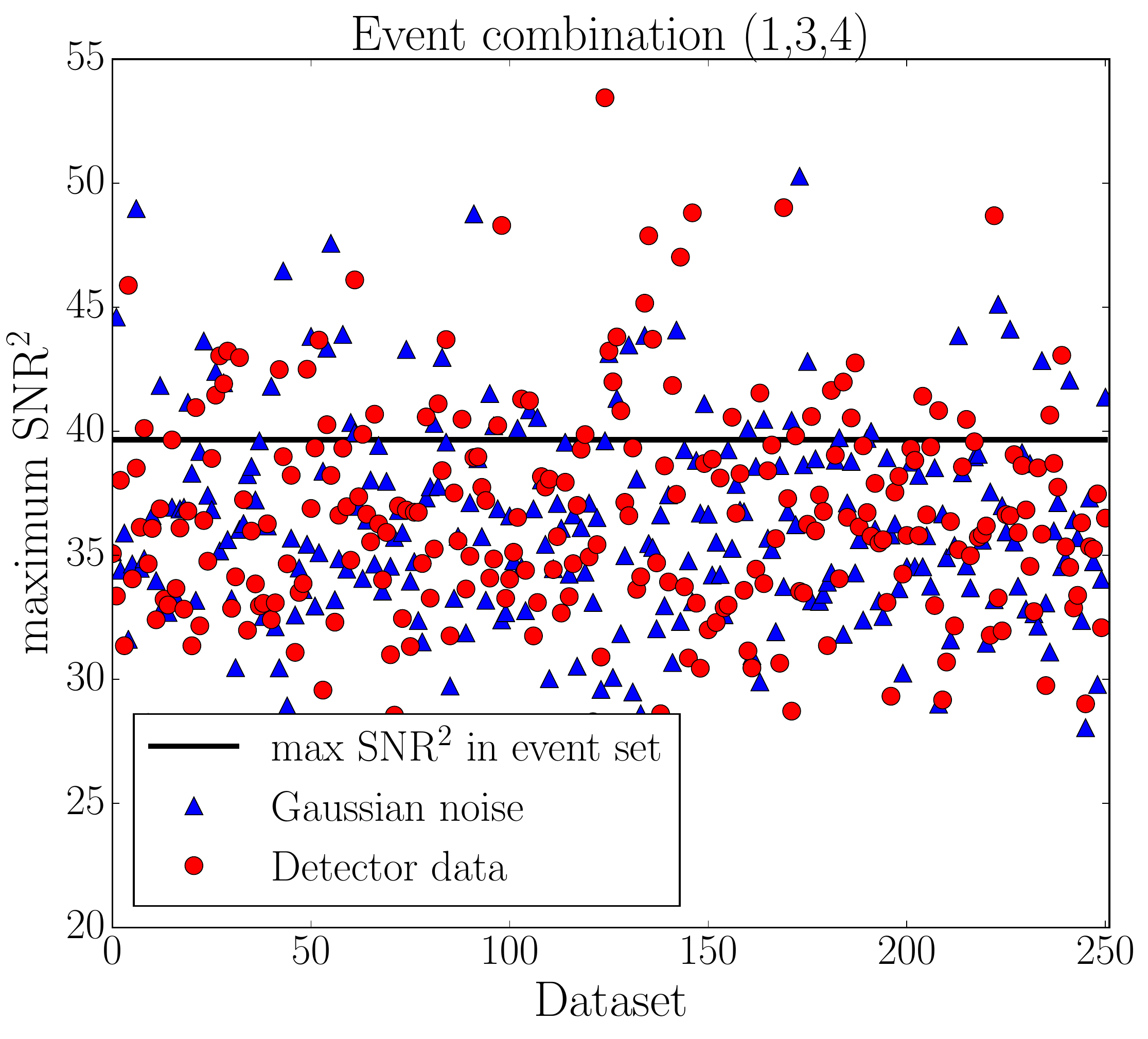}
    \caption{4096 seconds of LOSC data are divided into $\sim$ 256 segments of 16 seconds. The segments containing the GW event are excluded. For comparison, we generated 4096 seconds of Gaussian noise with the same PSD as estimated from the LOSC data for each event and divide it into segments in the same way. For each segment, the maximum SNR$^2$ for the echo search is shown. The distribution of SNR peaks in the data is similar to that in Gaussian noise.
	     The p-value is calculated from the number of points for detector data lying above the black line
	         which indicates the combined SNR value found immediately after each GW event.
	     There is no obvious trend in the peak SNR over time. 
            }
    \label{fig:distributionSNR}
\end{figure}

The combined background is shown in Fig. \ref{fig:distributionSNR} which shows the peak value of SNR$^2$ found in each segment for both the real detector data and Gaussian noise. For each event, the Gaussian noise was created with the same PSD as estimated from the data of this event. There is no significant difference between the distribution of peaks for detector data and for Gaussian noise. 
Here we note that there is no obvious trend in the peak SNR over time.
Also by this measure, there is no indication of the noise being unstable and preventing its use for background estimation. 
These two properties are shared by all single events and the alternate combination $(1,2,3)$: all show the similarity of the peak distribution for Gaussian noise and detector data, and lack a trend in time. 

\begin{table}
  \begin{centering}
    \begin{tabular}{c|c|c|c}
      Event & \cite{Abedi:2017v2} & original 16s (32s)& widened priors 16s (32s)\\
      \hline 
      GW150914 & 0.11 & 0.199 (0.238) & 0.705 (0.365) \\
      \hline
      LVT151012 & - & 0.056 (0.063) & 0.124 \\
      \hline
      GW151226 & - & 0.414 (0.476) & 0.837 \\
      \hline
      GW170104 & - & 0.725 & 0.757 \\
      \hline
      (1,2) & - & 0.004 & 0.36 \\
      \hline
      (1,3) & - & 0.159 & 0.801 \\
      \hline
      (1,2,3) & 0.011 & 0.020 (0.032) & 0.18 (0.144) \\
      \hline
      (1,3,4) & - & 0.199 (0.072) & 0.9 (0.32) \\
      \hline
      (1,2,3,4) & - & 0.044 (0.032) & 0.368 (0.112) \\
    \end{tabular}
    \caption{p-values obtained by using 4096 seconds of LOSC data divided into segments of 16 or 32 seconds length.
				Results are given for the priors in $t_0$ and $\gamma$
				    chosen in the original analysis and for widened priors,
					for the 3 O1 events individually
					and for various combinations of events.
				For the combinations directly comparable to \cite{Abedi:2017v2},
				    with the original priors,
				    we also record the Poisson errors
				    (as suggested in \cite{Abedi:2018pst}):
				    for GW150914 our p-values are $0.199\pm 0.028$ ($0.238\pm 0.043$),
				    and for (1,2,3) our p-values are $0.02\pm 0.009$ ($0.032\pm 0.016$).
				The Poisson errors for the full combination (1,2,3,4) with original priors,
				    are $0.044 \pm 0.013$ ($0.032 \pm 0.016$).
				With widened priors, the p-values are all much larger,
				    and the Poisson error relatively insignificant.
				Combinations that include LVT151012 have the lowest p-values.
				The addition of GW170104 to the three O1 events increases the combined p-value
				    and is thus more compatible with pure noise.
				The lowest p-value out of all 11 combinations using up to four events
				    is found for the combination (1,2). 
				Note however that considering more combinations of events
				    using the same data also leads to a higher effective trials factor to be accounted for.
}
    \label{tab:16+32_original+widened}
  \end{centering}
\end{table}

\begin{figure}
  \includegraphics[width=\columnwidth]{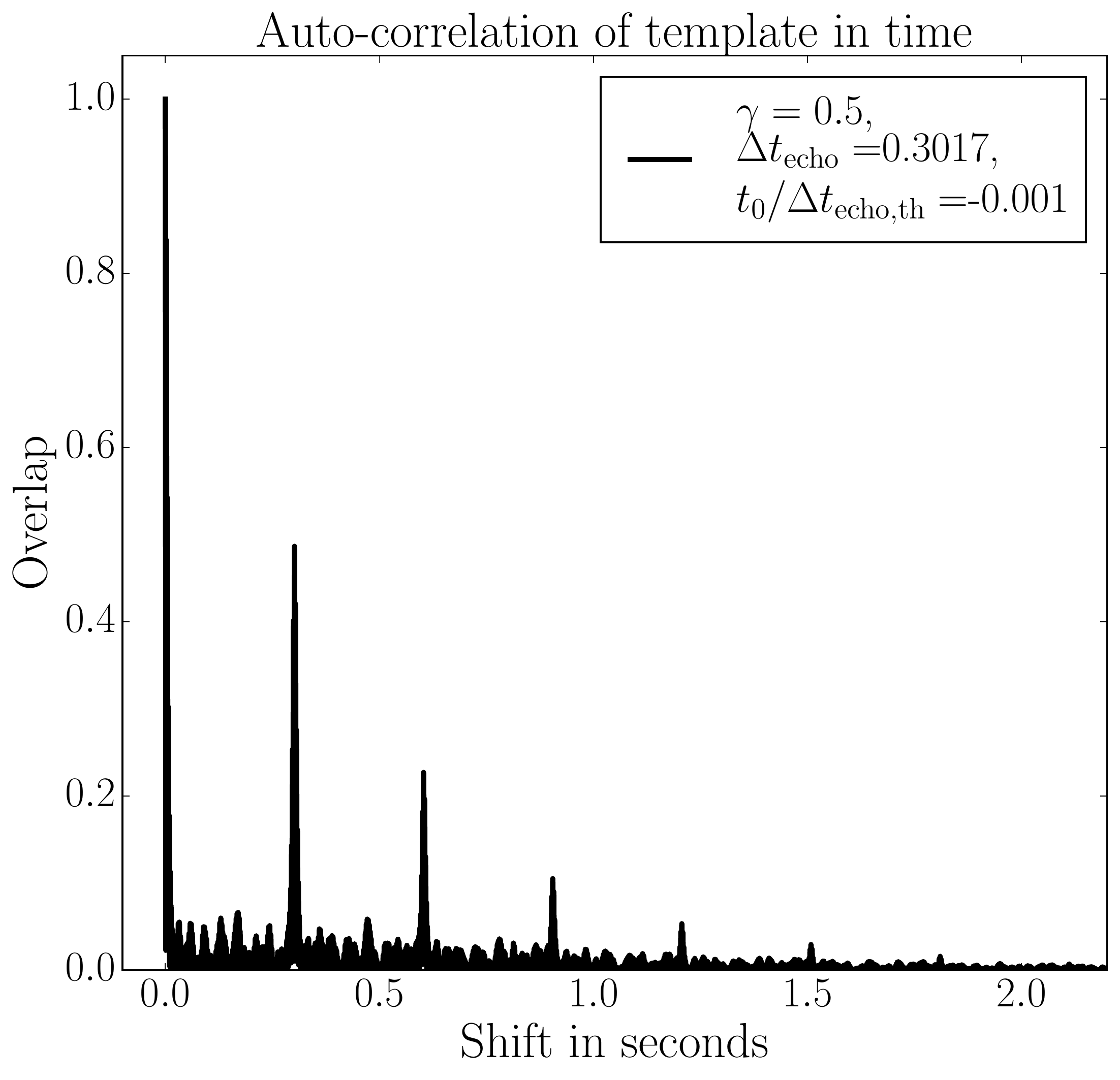}
  \caption{Autocorrelation of the pure echo template depending on shift in time.
            The peaks demonstrate that there is significant correlation between background samples for small shifts.}
  \label{fig:overlap}
\end{figure}

A second concern about the background estimation used in \cite{Abedi:2016hgu, Abedi:2017isz, Abedi:2017v2}
	arises from the very small shift in time between samples that are considered independent.
In this, the quasi-periodic nature of the echo signal has to be considered,
	leading to potentially long templates with equally quasi-periodic autocorrelation in time.
The autocorrelation of the echo templates is shown in Fig. \ref{fig:overlap}.
This affects the significance estimation as performed in \cite{Abedi:2016hgu,Abedi:2017v2}. 
This method is problematic because the template is significantly longer than the shift in time between background sample intervals.
If the autocorrelation between templates used in different background samples does not vanish,
    the results from these samples cannot be considered truly independent:
    a feature of the data at one point in time then influences the SNR found for several background samples. 
The total number of effectively independent samples is thus much lower in this method.

    As shown in Fig. \ref{fig:overlap}, 
the echo signal model leads to a series of peaks in the overlap of original and time-shifted wavefunction,
    depending on the parameters $\gamma$, $\Delta t_{\rm echo}$ and $t_0$.
A value of overlap to be considered sufficiently independent could be e.g. $1 \%$. 
To achieve this value,
    there are two ways to place templates with respect to the original position in time.
For $\gamma < 1$, i.e. a damped echo signal,
	applying a shift in time by a sufficiently large multiple of $\Delta t_{\rm echo}$ leads to a reduction of the correlation.
Using the GW150914 template and $\gamma = 0.5$ shows that at least $7$ times $\Delta t_{\rm echo}$ is necessary.
In Fig. \ref{fig:overlap}, this corresponds to the very small peak close to $2.1$ seconds of time-shift. 

Alternatively, the templates can be interlaced, such that the echoes of one template are placed within the time between echoes of the other.
This corresponds to the small overlap values between the peaks in Fig. \ref{fig:overlap}.
Here we again use the GW150914 template and the most favourable values in the prior range,
	i.e. the shortest echoes ($t_0 = -0.001 \Delta t_{\text{echo, theory}}$) and the longest delay ($\Delta t_{\rm echo} = 0.30166 \: {\rm s}$). 
Then about $7$ echoes can be placed between the peaks of the original template.
For these parameters we now consider shifts in time of the echo template up to $29 \Delta t_{\rm echo}$,
    which is the maximum shift used in ADA's significance estimation. 
We find about $4$ independent samples through sufficient timeshift and a factor $8$ through interlacing,
	giving $\sim 32$ independent samples. 

As the maximisation is performed over a range of parameters,
	exactly determining the total number of independent samples would need further consideration.
The parameters chosen for this estimate, however, are favourable,
	as smaller damping or smaller time delay would further lower the total number.
For the maximisation combining the different events,
	$\Delta t_{\rm echo}$ may vary independently between events,
	obstructing a clear estimate on the number of samples;
	the same considerations, however, still apply. 
These considerations suggest the method of \cite{Abedi:2017v2} contains only a small number of independent background samples, on the order of a few tens of samples.

The method we employed to estimate the background precludes this effect
	by only applying the matched filtering procedure to separate sets of data.
The template thus is always placed in only one of the background samples and the resulting SNR cannot be influenced by data features in the remaining samples. 

The nature of the echo templates leads to a further potential problem: due to the delay between echoes, low frequency components may be introduced in the template.
Due to the delay times of about 0.1 to 0.3 seconds, these frequencies are expected to be in the range below 20 Hz, down to a few Hz.
However, the data as supplied by the LOSC, \cite{LOSC}, is not calibrated below 10 Hz, as mentioned in the corresponding notes on data usage.
The results of the analysis may thus be influenced by the uncalibrated data.
We have repeated the analysis after applying a high-pass filter to the data and the original waveform, removing the data below 30 Hz for final SNR calculations.
The results of this analysis are almost identical to those before applying the high-pass filter in terms of SNR.
The resulting p-values similarly show only minor changes compared to the values given in Table \ref{tab:16+32_original+widened}.
The combination of introduced low frequency components in the templates and the uncalibrated data thus seems to have no significant effect on the results of the analysis.

\section{Remark on echo templates}

\begin{figure}
  \includegraphics[width=\columnwidth]{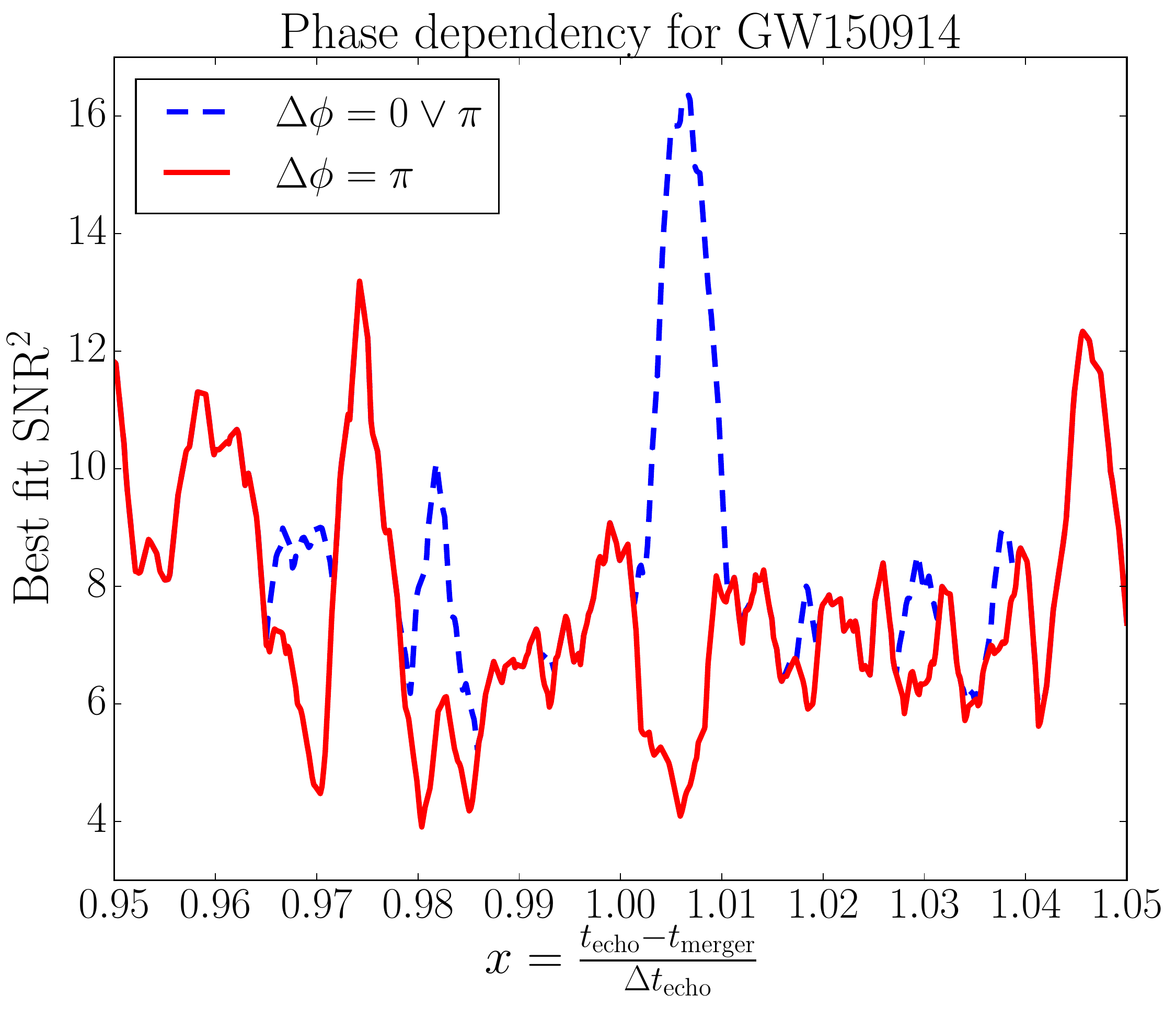}
  \caption{The maximised SNR for GW150914 is shown with different phase change $\Delta \phi$ between event and first echo.
    The original analysis allows for $\Delta \phi = 0 \lor \pi$ for each $x$ and then finds a prominent peak near $x=1$.
    However the model requires $\Delta \phi = \pi$.
    When we enforce this value in the analysis, the peak vanishes. }
  \label{fig:GW150914_phaseflip}
\end{figure}

The \textbf{ADA-search} procedure does not distinguish between inversion and non-inversion of the first echo's phase. 
The waveform templates used here are based on the simple model described in \cite{Abedi:2016hgu,Abedi:2017v2}.
Within this model, the phase-change of the gravitational wave between the original signal and the first echo is described as a simple phase-flip. However, as only the square of the SNR from the matched filter analysis is considered for the maximisation,
	the result is insensitive to this phase inversion.
Repeating the analysis for GW150914 and enforcing the phase inversion as required by this model,
	we find that the prominent peak in SNR at $x = 1$ vanishes.
This is shown in Fig. \ref{fig:GW150914_phaseflip}.
More sophisticated models would be needed to determine whether this phase flip is truly required or not. 
Still, it is worth noting that the peak that forms the basis of evidence for echoes in \cite{Abedi:2016hgu,Abedi:2017v2},
	does not contain this phase inversion as required by the simple model, but actually the opposite phase. 
\newline 
\section{Conclusions}
A full analysis of the data at a level necessary to confidently detect echo signals is outside the scope of this work.
However, we have analysed the data using a simple templated search similar to \cite{Abedi:2016hgu}.
Using an extended background estimated from the full 4096 seconds of data released publicly by the LIGO collaboration for each event in the first observing run, we find a p-value for the null noise only hypothesis of $0.02$, higher than that reported using the restricted background of $0.011$ in \cite{Abedi:2016hgu}. 
We have demonstrated the importance of LVT151012, the weakest LIGO candidate event \cite{DetChar}, in obtaining this p-value.
A combined analysis of the three events GW150914, GW151226 and GW170104, excluding LVT151012, yields an even larger p-value of $0.199$, fully consistent with noise.
We have also identified a number of weaknesses in the analysis method of \cite{Abedi:2016hgu} including the role of the prior boundaries and the density of templates.
In particular we have examined the role of the $\gamma$ parameter and found that the clustering of $\gamma$ values near $\gamma = 1$ is entirely consistent with noise.
The signal manifold is such that in pure Gaussian noise, one would expect many more triggers with values of $\gamma \sim 1$.
This perhaps would not be expected for true echo signals, although a more detailed model of echo signals would be needed to make a quantitative prediction.
A similar bias in recovered parameters concerns the peak amplitude,
    which for both GW150914 and LVT151012 was found by \cite{Abedi:2016hgu,Abedi:2017v2} just on the boundary of credible signal recovery.

In conclusion, we find that the tentative evidence as presented
	in \cite{Abedi:2016hgu, Abedi:2017v2, Abedi:2017isz} is lacking in several key aspects with respect to template placement, background estimation and implementation.
Our analysis of these short shortcomings shows that the method of Abedi et al. cannot
	provide observational evidence for or against the existence of near-horizon
	Planck-scale structure in black holes.
This stresses the importance of developing both new theoretical models and analysis methods for gravitational wave echoes from such structures. 
We hope some of the concerns explored here may be useful to vet other searches for echoes,
	such as \cite{Conklin:2017lwb},
	and help in the development of methods which would place
	black hole near-horizon physics within the realm of gravitational wave observations.

\begin{acknowledgments}
The authors thank Andrew Lundgren, Laura Nuttall, Vitor Cardoso,
	and the authors of \cite{Abedi:2016hgu, Abedi:2017v2, Abedi:2017isz}, 
	for useful discussions, as well as Bruce Allen for helpful comments. 
Some of the discussions particularly enjoyed the hospitality of
	meetings at Nikhef and at the Perimeter Institute.
This research has made use of data,
software and/or web tools obtained from the LIGO Open Science Center (https://losc.ligo.org),
a service of LIGO Laboratory and the LIGO Scientific Collaboration.
LIGO is funded by the U.S. National Science Foundation (NSF).
O.B. acknowledges the NSF for financial support from Grant No. PHY-1607520.
\end{acknowledgments}


\begin{thebibliography}{99}

\bibitem{Abbott:2016blz}
  \CITELVC,
  \LONGBIB{``Observation of Gravitational Waves from a Binary Black Hole Merger,''}{}
  Phys.\ Rev.\ Lett.\  {\bf 116} (2016) no.6,  061102
  \CITEDOI{10.1103/PhysRevLett.116.061102}
  [arXiv:1602.03837\ARXIVFULL{gr-qc}].

\bibitem{Abbott:2016nmj}
  \CITELVC,
  \LONGBIB{``GW151226: Observation of Gravitational Waves from a 22-Solar-Mass Binary Black Hole Coalescence,''}{}
  Phys.\ Rev.\ Lett.\  {\bf 116} (2016) no.24,  241103
  \CITEDOI{10.1103/PhysRevLett.116.241103}
  [arXiv:1606.04855\ARXIVFULL{gr-qc}].

\bibitem{TheLIGOScientific:2016pea}
  \CITELVC,
  \LONGBIB{``Binary Black Hole Mergers in the first Advanced LIGO Observing Run,''}{}
  Phys.\ Rev.\ X {\bf 6} (2016) no.4,  041015
  \CITEDOI{10.1103/PhysRevX.6.041015}
  [arXiv:1606.04856\ARXIVFULL{gr-qc}].

\bibitem{Abbott:2017vtc}
  \CITELVC,
  \LONGBIB{``GW170104: Observation of a 50-Solar-Mass Binary Black Hole Coalescence at Redshift 0.2,''}{}
  Phys.\ Rev.\ Lett.\  {\bf 118}, no. 22, 221101 (2017)
  \CITEDOI{doi:10.1103/PhysRevLett.118.221101}
  [arXiv:1706.01812\ARXIVFULL{gr-qc}].

\bibitem{Abbott:2017gyy}
  \CITELVC,
  \LONGBIB{``GW170608: Observation of a 19-solar-mass Binary Black Hole Coalescence,''}{}
  Astrophys.\ J.\  {\bf 851}, no. 2, L35 (2017)
  doi:10.3847/2041-8213/aa9f0c
  [arXiv:1711.05578 [astro-ph.HE]].

\bibitem{Abbott:2017oio}
  \CITELVC,
  \LONGBIB{``GW170814: A Three-Detector Observation of Gravitational Waves from a Binary Black Hole Coalescence,''}{}
  Phys.\ Rev.\ Lett.\  {\bf 119}, no. 14, 141101 (2017)
  doi:10.1103/PhysRevLett.119.141101
  [arXiv:1709.09660\ARXIVFULL{gr-qc}].

\bibitem{TheLIGOScientific:2016wfe}
  \CITELVC,
  \LONGBIB{``Properties of the Binary Black Hole Merger GW150914,''}{}
  Phys.\ Rev.\ Lett.\  {\bf 116} (2016) no.24,  241102
  \CITEDOI{10.1103/PhysRevLett.116.241102}
  [arXiv:1602.03840\ARXIVFULL{gr-qc}].
 
\bibitem{TheLIGOScientific:2016src}
  \CITELVC,
  \LONGBIB{``Tests of general relativity with GW150914,''}{}
  Phys.\ Rev.\ Lett.\  {\bf 116} (2016) no.22,  221101
  \CITEDOI{10.1103/PhysRevLett.116.221101}
  [arXiv:1602.03841\ARXIVFULL{gr-qc}].

\bibitem{MembraneParadigm}
  D.~A.~MacDonald, R.~H.~Price, K.~S.~Thorne(editors)
  \LONGBIB{``Black Holes: The Membrane Paradigm''}{}
  Yale University Press (1986)
  
\bibitem{Almheiri:2012rt}
  A.~Almheiri, D.~Marolf, J.~Polchinski and J.~Sully,
  \LONGBIB{``Black Holes: Complementarity or Firewalls?,''}{}
  JHEP {\bf 1302} (2013) 062
  \CITEDOI{10.1007/JHEP02(2013)062}
  [arXiv:1207.3123\ARXIVFULL{hep-th}].
  
\bibitem{Harry:2010zz}
  G.~M.~Harry [LIGO Scientific Collaboration],
  \LONGBIB{``Advanced LIGO: The next generation of gravitational wave detectors,''}{}
  Class.\ Quant.\ Grav.\  {\bf 27}, 084006 (2010).
  doi:10.1088/0264-9381/27/8/084006

\bibitem{TheLIGOScientific:2014jea} 
  J.~Aasi {\it et al.} [LIGO Scientific Collaboration],
  \LONGBIB{``Advanced LIGO,''}{}
  Class.\ Quant.\ Grav.\  {\bf 32}, 074001 (2015)
  doi:10.1088/0264-9381/32/7/074001
  [arXiv:1411.4547 [gr-qc]].

\bibitem{TheVirgo:2014hva} 
  F.~Acernese {\it et al.} [VIRGO Collaboration],
  \LONGBIB{``Advanced Virgo: a second-generation interferometric gravitational wave detector,''}{}
  Class.\ Quant.\ Grav.\  {\bf 32}, no. 2, 024001 (2015)
  doi:10.1088/0264-9381/32/2/024001
  [arXiv:1408.3978 [gr-qc]].

\bibitem{Cardoso:2016rao}
  V.~Cardoso\LONGBIB{, E.~Franzin \AUTHAND P.~Pani}{{~\it et al.}},
  \LONGBIB{``Is the gravitational-wave ringdown a probe of the event horizon?,''}{}
  Phys.\ Rev.\ Lett.\  {\bf 116} (2016) no.17,  171101
   Erratum: [Phys.\ Rev.\ Lett.\  {\bf 117} (2016) no.8,  089902]
  \CITEDOI{10.1103/PhysRevLett.117.089902, 10.1103/PhysRevLett.116.171101}
  [arXiv:1602.07309\ARXIVFULL{gr-qc}].

  
  
\bibitem{Cardoso:2016oxy}
  V.~Cardoso\LONGBIB{, S.~Hopper, C.~F.~B.~Macedo, C.~Palenzuela \AUTHAND P.~Pani}{{~\it et al.}},
  \LONGBIB{``Gravitational-wave signatures of exotic compact objects and of quantum corrections at the horizon scale,''}{}
  Phys.\ Rev.\ D {\bf 94} (2016) no.8,  084031
  \CITEDOI{10.1103/PhysRevD.94.084031}
  [arXiv:1608.08637\ARXIVFULL{gr-qc}].
 
\bibitem{Cardoso:2017njb} 
  V.~Cardoso\LONGBIB{ \AUTHAND P.~Pani}{{~\it et al.}},
  \LONGBIB{``The observational evidence for horizons: from echoes to precision gravitational-wave physics,''}{}
  [arXiv:1707.03021 [gr-qc]].

\bibitem{Mark:2017dnq} 
  Z.~Mark\LONGBIB{, A.~Zimmerman, S.~M.~Du \AUTHAND Y.~Chen}{{~\it et al.}},
  \LONGBIB{``A recipe for echoes from exotic compact objects,''}{}
  Phys.\ Rev.\ D {\bf 96} (2017) no.8,  084002
  \CITEDOI{10.1103/PhysRevD.96.084002}
  [arXiv:1706.06155 [gr-qc]].

\bibitem{Volkel:2017kfj} 
  S.~H.~V\"{o}lkel\LONGBIB{ \AUTHAND K.~D.~Kokkotas}{{~\it et al.}},
  \LONGBIB{``Ultra Compact Stars: Reconstructing the Perturbation Potential,''}{}
  Class.\ Quant.\ Grav.\  {\bf 34} (2017) no.17,  175015
  \CITEDOI{10.1088/1361-6382/aa82de}
  [arXiv:1704.07517 [gr-qc]].

\bibitem{Cardoso:2017cfl} 
  V.~Cardoso\LONGBIB{, E.~Franzin, A.~Maselli, P.~Pani \AUTHAND G.~Raposo}{{~\it et al.}},
  \LONGBIB{``Testing strong-field gravity with tidal Love numbers,''}{}
  Phys.\ Rev.\ D {\bf 95}, no. 8, 084014 (2017)
  Addendum: [Phys.\ Rev.\ D {\bf 95}, no. 8, 089901 (2017)]
  doi:10.1103/PhysRevD.95.089901, 10.1103/PhysRevD.95.084014
  [arXiv:1701.01116 [gr-qc]].

\bibitem{Abedi:2016hgu}
  J.~Abedi\LONGBIB{, H.~Dykaar \AUTHAND N.~Afshordi}{{~\it et al.}},
  \LONGBIB{``Echoes from the Abyss: Evidence for Planck-scale structure at black hole horizons,''}{}
  [arXiv:1612.00266v1\ARXIVFULL{gr-qc}],
	1 Dec 2016.
  
\bibitem{Abedi:2017v2}
  J.~Abedi\LONGBIB{, H.~Dykaar \AUTHAND N.~Afshordi}{{~\it et al.}},
  \LONGBIB{``Echoes from the Abyss: Tentative evidence for Planck-scale structure at black hole horizons,''}{}
  Phys.\ Rev.\ D {\bf 96}, no. 8, 082004 (2017)
  doi:10.1103/PhysRevD.96.082004
  [arXiv:1612.00266v2\ARXIVFULL{gr-qc}],
	3 Oct 2017.

\bibitem{Abedi:2017isz}
  J.~Abedi\LONGBIB{, H.~Dykaar \AUTHAND N.~Afshordi}{{~\it et al.}},
  \LONGBIB{``Echoes from the Abyss: The Holiday Edition!,''}{}
  [arXiv:1701.03485\ARXIVFULL{gr-qc}].

 \bibitem{LOSC}
  LIGO Scientific Collaboration,
  ``LIGO Open Science Center - Data Releases for Observed Transients",
  2017,
	\CITEDOI{10.7935/K5MW2F23, 10.7935/K5CC0XMZ, 10.7935/K5H41PBP, 10.7935/K53X84K2}
  \url{http://losc.ligo.org/events}

\bibitem{Vallisneri:2014vxa} 
  M.~Vallisneri\LONGBIB{, J.~Kanner, R.~Williams, A.~Weinstein \AUTHAND  B.~Stephens}{{~\it et al.}},
  \LONGBIB{``The LIGO Open Science Center,''}{}
  J.\ Phys.\ Conf.\ Ser.\  {\bf 610}, no. 1, 012021 (2015)
  doi:10.1088/1742-6596/610/1/012021
  [arXiv:1410.4839 [gr-qc]].

\bibitem{Nakamura:2016gri} 
  T.~Nakamura\LONGBIB{, H.~Nakano \AUTHAND T.~Tanaka}{{~\it et al.}},
  \LONGBIB{``Detecting quasinormal modes of binary black hole mergers with second-generation gravitational-wave detectors,''}{}
  Phys.\ Rev.\ D {\bf 93} (2016) no. 4, 044048
  \CITEDOI{10.1103/PhysRevD.93.044048}
  [arXiv:1601.00356\ARXIVFULL{astro-ph.HE}].

\bibitem{Holdom:2016nek} 
  B.~Holdom\LONGBIB{ \AUTHAND J.~Ren}{{~\it et al.}},
  \LONGBIB{``Not quite a black hole,''}{}
  Phys.\ Rev.\ D {\bf 95} (2017) no.8,  084034
  \CITEDOI{10.1103/PhysRevD.95.084034}
  [arXiv:1612.04889\ARXIVFULL{gr-qc}].

\bibitem{Connaughton:2016umz}
  V.~Connaughton {\it et al.},
  \LONGBIB{``Fermi GBM Observations of LIGO Gravitational Wave event GW150914,''}{}
  Astrophys.\ J.\  {\bf 826} (2016) no.1,  L6
  \CITEDOI{10.3847/2041-8205/826/1/L6}
  [arXiv:1602.03920\ARXIVFULL{astro-ph.HE}],
  Appendix B, `Significance of two-parameter coincidence'.

\bibitem{EchoComments} 
  G.~Ashton\LONGBIB{, O.~Birnholtz, M.~Cabero, C.~Capano, T.~Dent, B.~Krishnan, G.~D.~Meadors, A.~B.~Nielsen, A.~Nitz \AUTHAND J.~Westerweck}{{~\it et al.}},
  \LONGBIB{``Comments on: "Echoes from the abyss: Evidence for Planck-scale structure at black hole horizons",''}{}
  [arXiv:1612.05625\ARXIVFULL{gr-qc}].

\bibitem{pvalues}
  R.~L.~Wasserstein and N.~A.~Lazar,
  \LONGBIB{``The ASA's Statement on p-Values: Context, Process, and Purpose,''}{}
  The American Statistician, {\bf 70:2} (2016), 129-133
  \CITEDOI{10.1080/00031305.2016.1154108}
  \url{https://www.amstat.org/asa/files/pdfs/P-ValueStatement.pdf}

\bibitem{AbediPrivate}
 J.~Abedi,
  Private communication, applying the method of \cite{Abedi:2017isz} to the posteriors of \cite{Abbott:2017vtc}, yielding $\Delta t_{\rm echo} = 0.2412 \pm 0.018449$ s.

\bibitem{Abedi:2018pst}
  J.~Abedi, H.~Dykaar and N.~Afshordi,
  ``Comment on: "Low significance of evidence for black hole echoes in gravitational wave data",''
  arXiv:1803.08565 [gr-qc].

\bibitem{LOSCtutorial}
  LIGO Scientific Collaboration,
  ``LOSC Event tutorial v1.4",
  2016,
  \url{http://losc.ligo.org/s/events/GW150914/LOSC_Event_tutorial_GW150914.html}

\bibitem{DetChar}
  \CITELVC,
  \LONGBIB{``Characterization of transient noise in Advanced LIGO relevant to gravitational wave signal GW150914,''}{}
  Class.\ Quant.\ Grav.\  {\bf 33}, no. 13, 134001 (2016)
  \CITEDOI{10.1088/0264-9381/33/13/134001}
  [arXiv:1602.03844\ARXIVFULL{gr-qc}].

\bibitem{Conklin:2017lwb} 
  R.~S.~Conklin\LONGBIB{, B.~Holdom \AUTHAND J.~Ren}{{~\it et al.}},
  \LONGBIB{``Gravitational wave echoes through new windows",''}{}
  [arXiv:1712.06517\ARXIVFULL{gr-qc}].

\end{thebibliography}
\end{document}